\documentclass[a4paper,11pt]{article}

\usepackage{geometry}
\usepackage{fancyhdr}
\usepackage{amsmath, amssymb}
\usepackage{graphicx}
\usepackage{color}
\definecolor{bg}{rgb}{0.89, 0.95, 0.71} 
\definecolor{ac}{rgb}{0.1, 0.1, 0.9} 
\definecolor{rc}{rgb}{0.77, 0.57, 0.71} 
\usepackage{listings}
\lstdefinestyle{latex}{language=TeX,
                       backgroundcolor=\color{bg},
                       basicstyle=\small\ttfamily,
                       frame=leftline,
                       xleftmargin=1.4em,
                       framexleftmargin=.8em}
\lstdefinestyle{cmdline}{
                         }
\usepackage{url}
\usepackage[pdftex,
            bookmarks,
            colorlinks=false,
            linkcolor=black,
            plainpages = false,
            pdfpagemode = UseNone,
            pdfstartview = FitH,
            citecolor = green, urlcolor = cyan, filecolor = magenta]{hyperref}

\usepackage[para]{footmisc}
\makeatletter
\long\def\@makefntext#1{\leavevmode
\@makefnmark\nobreak
\hskip.05em\relax#1%
}
\makeatother

\newcommand{\bibliolink}[2]{\href{#1}{\nolinkurl{#1}}.}

\makeatletter 

\def\section{\@startsection{section}{1}{0pt}{2.5ex plus .6ex minus
    .2ex}{1.0ex plus .15ex}{\large\bf}} 
\def\subsection{\@startsection{subsection}{2}{0pt}{1.5ex plus .3ex minus
   .1ex}{.2ex plus .1ex}{\normalsize\bf}} 
\makeatother

\newlength{\rulelength}
\makeatletter
\renewcommand{\@makecaption}[2]{%
		    \sbox{\@tempboxa}{{\small #1: #2}}
		    \ifdim \wd\@tempboxa > \hsize
		    {\small #1: #2}\par
		    \else
		    \global\@minipagefalse
		    \hbox to \hsize {\hfil {\small #1: #2}\hfil}%
		    \fi
		   \vspace{\belowcaptionskip}
		     }
\makeatother

\begin{document}

\begin{center}
\thispagestyle{empty}

{\bf\large Spectra and Mass Composition of Ultrahigh--Energy Cosmic Rays\\
from Point Sources}\\[4mm]
Nikolai N. Kalmykov\footnote{\nolinkurl{kalm@eas.sinp.msu.ru}}\\[1mm]
{\it\small Skobeltsyn Institute of Nuclear Physics, Lomonosov Moscow State University,\\Leninskie gory 1/2, Moscow, Russia}\\[4mm]
Olga P. Shustova\footnote{\nolinkurl{olga.shustova@eas.sinp.msu.ru}}\\[1mm]
{\it\small Faculty of Physics, Lomonosov Moscow State University,\\Leninskie gory 1/2, Moscow, Russia}\\[4mm]
Anna V. Uryson\footnote{\nolinkurl{uryson@sci.lebedev.ru}}\\[1mm]
{\it\small Lebedev Physical Institute, Russian Academy of Sciences,\\Leninsky prospect 53, Moscow, Russia}
\end{center}

\begin{abstract}
We present spectra and mass composition of cosmic rays incoming to the Earth in the energy range $(0.5\!-\!2)\!\cdot\!10^{20}$\,eV. As their sources we consider Seyfert galaxies located at distances $\lesssim\!40$~Mpc, following an acceleration model for such moderate--power objects. Mass composition of the particles at sources is assumed to be mixed. Generation spectra are described by a function $E^{-\gamma_0}$, where $\gamma_0$ is an arbitrary parameter. It is shown that the assumptions adopted make it possible to describe experimental data provided by HiRes and Pierre Auger Observatory, using different values of $\gamma_0$.
\end{abstract}

\section{Introduction}\label{intr}
Since the Greisen--Zatsepin--Kuzmin (GZK) cutoff for protons with energies $\gtrsim\!5\!\cdot\!10^{19}$\,eV was predicted in 1966 \cite{GZKI,GZKII}, issues regarding ultrahigh--energy cosmic rays (UHECRs) have remained one of the main topics in high--energy astrophysics. Because of vanishingly small flux of such energetic particles, arrays enormous in area should be used over a~long period of~time. At present a~sufficient number of experimental events have been collected to provide us with an~idea on the behavior of the differential energy spectrum of UHECRs. Now the existence of the GZK cutoff is generally considered to be proved by measurements of the largest UHECR facilities, HiRes \cite{HiRes_GZK} and Pierre Auger Observatory (Auger) \cite{PAO_GZK}. It is appropriate, however, to recall that the experimental data reported by AGASA \cite{AGASA_GZK} indicated the~opposite effect.

Although the suppression of the CR flux at ultrahigh energies has been reported by both HiRes and Auger, their spectra differ significantly in shape. For instance, while the Hires collaboration clearly locates the start point of the GZK cutoff \cite{Sokolsky_2009}, one of two fits of the Auger spectrum has a~smooth form \cite{PAO_2009}. Below we discuss the approximations suggested by both teams. Another, more surprising, contradiction between their observations is that the HiRes results demonstrate the proton--dominated CR composition at energies from $2\!\cdot\!10^{18}$\,eV up to $5\!\cdot\!10^{19}$\,eV \cite{HiRes_xmax}, whereas Auger points out a~transition from protons to heavy nuclei, starting with energy of a~few times $10^{18}$\,eV \cite{Auger_xmax}. This problem was discussed, for example, in~\cite{PhysTod}.

To describe experimental data and therefore to determine likely sources of UHECRs, various models of initial mass composition and acceleration conditions were considered (see, for instance, \cite{Harari_2006}\,--\,\cite{Aloisio_2009}). Nevertheless the origin of ultrahigh--energy particles remains an~open question. In this paper we present spectra and mass composition obtained within a~model \cite{Uryson_2001,Uryson_2004} in which nuclei of moderate--power Seyfert galaxies are put forward as UHECR sources. Comparison of calculation results with measurements allows us to reveal appropriate acceleration regimes and besides to define somehow distances to the most probable sources. It is well known that particles generated above the GZK threshold shift to the region of lower energies in propagating through the space, shaping peculiarities in the cosmic ray spectrum, the so-called dip and bump \cite{Hill_1985}\,--\,\cite{Berezinsky_2006}. Study of these spectral features provides insight into UHECR sources. But a~more natural approach for investigation of UHECR particles is to exploit data near their maximum acceleration energy, so here we restrict our consideration to the region above the GZK cutoff.

\section{Assumptions}\label{assump}
Main features of the above--mentioned model are as follows. Charged particles are accelerated to ultrahigh energies at the front of shocks generated in relativistic jets of sources. The jet matter resembles one of the accretion disk, therefore the accelerated particles can be both protons and heavy ions. Because of lack of information about magnetic field strength within the jets, its value can be regarded as an~arbitrary parameter specified on the assumption that particles should reach their maximum acceleration energies. (Though magnetic field within the jets of some active galactic nuclei has been estimated, e.g.~\cite{Artykh,McCann}.) The expression of the maximum energies for nuclei with charge $Z\!\geqslant\!2$, derived on basis of this model, has the form:
\begin{equation}\label{maxen_nucl}
E_\mathrm{max}=9.3\cdot10^{19}Z^{2/3}\;\text{eV}.
\end{equation}
The dependence $Z^{2/3}$ arises due to energy losses (see \cite{Uryson_2001,Uryson_2004}). For protons it is necessary to take into account $A\!=\!Z\!=\!1$ so their maximum energy is equal to
\begin{equation}\label{maxen_prot}
E_\mathrm{max}^\mathrm{p}\approx4\cdot10^{19}\;\text{eV}.
\end{equation}
In~Section \ref{res} the numerical values in Eq.\,\eqref{maxen_nucl} and \eqref{maxen_prot} are varied a~bit to obtain a~better agreement with the experimental data. Such operation is fully justified as at the moment it is hardly possible to set these values unambiguously.

The mass composition of the particles at sources is assumed to be mixed. We consider stable nuclei with the highest natural content. Besides protons, these are nuclei of $^4\mathrm{He}$, $^{12}\mathrm{C}$, $^{14}\mathrm{N}$, $^{16}\mathrm{O}$, $^{20}\mathrm{Ne}$, $^{23}\mathrm{Na}$, $^{24}\mathrm{Mg}$, $^{27}\mathrm{Al}$, $^{28}\mathrm{Si}$, $^{32}\mathrm{S}$, $^{40}\mathrm{Ar}$, $^{40}\mathrm{Ca}$, $^{52}\mathrm{Cr}$ and $^{56}\mathrm{Fe}$. Their abundance ratios are taken from~\cite{Allen}.

\begin{figure}[t]
\begin{center}
\includegraphics[height=14pc]{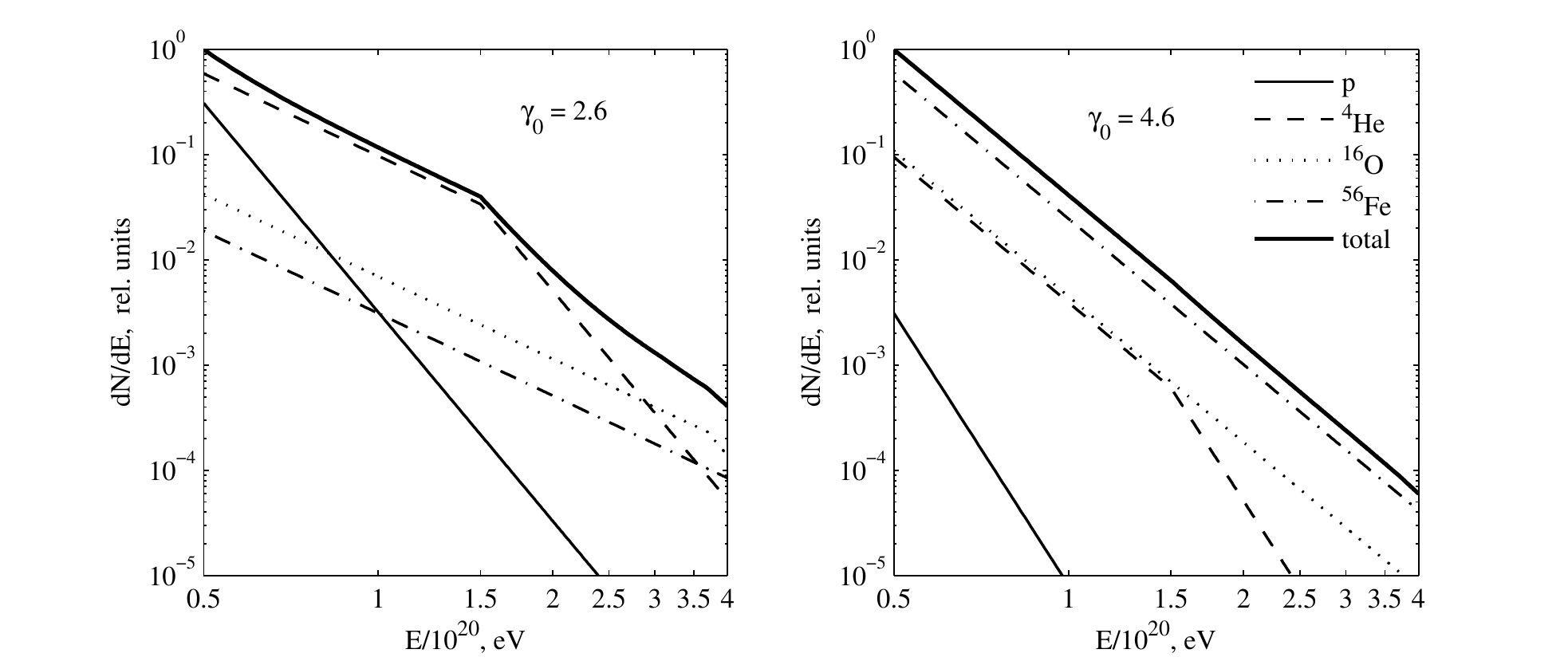}
\end{center}
\vspace{-0.5pc}
\caption{Differential energy spectra of some abundant nuclei ($\mathrm{p}$, $^4\mathrm{He}$, $^{16}\mathrm{O}$, $^{56}\mathrm{Fe}$) at a source, with $\gamma_0\!=\!2.6$ (the left panel) and $\gamma_0\!=\!4.6$ (the right panel). The total spectrum (solid thick curve) is normalized to unity at $E\!=\!5\!\cdot\!10^{19}$\,eV.}\label{init_spectra}
\end{figure}

The model proposed makes it possible to estimate the maximum acceleration energies $E_\mathrm{max}$ but does not specify the shape of the energy spectrum near them. So we adopt a~reasonable assumption that particle energy spectra at sources obey a~broken inverse power law with indices~$\gamma_0$ at energies $E\!\leqslant\!E_\mathrm{max}$ and $\gamma_0\!+\!4$ otherwise. A~similar form was used in \cite{Sveshnikova_2004} to describe roughly the energy spectrum of cosmic rays accelerated in supernova shocks. It is necessary to stress that at large values of~$\gamma_0$ the broken power law should be used only in the vicinity of $E_\mathrm{max}$ as its extrapolation to substantially lower energies may demand too high efficiency from the sources. In the case of a~power law spectrum the number of nuclei with mass number~$A$ and energy in the range of $(E,E\!+\!dE)$ is proportional to $A^{\gamma_0-1}$ \cite{Ginzburg}. If the injection index is high enough, the fraction of heavy nuclei is essentially increased at a~given energy. The generation spectra of some abundant nuclei at energies of $(0.5-4)\cdot10^{20}$\,eV are presented in Figure~\ref{init_spectra}, with two different values of the index. For~$\gamma_0\!=\!2.6$, nuclei of helium dominate among the particles up to $\approx\!2.5\!\cdot\!10^{20}$\,eV, while nuclei of oxygen and iron prevail alternately, starting with $E\!\approx\!3\!\cdot\!10^{20}$\,eV. If $\gamma_0\!=\!4.6$, the pattern changes substantially and iron nuclei begin to dominate over the whole energy range of~interest.

The acceleration model adopted was developed for moderate--power Seyfert nuclei. As possible sources of UHECRs, this type of active galactic nuclei (AGNs), with redshifts $z\!\lesssim\!0.009$, was identified in \cite{Uryson_1996,Uryson_AR}. We follow this hypothesis and refer to them as nearby sources. Their distances are set according to~\cite{catalog}.

Based on these assumptions we calculate spectra and mass composition of cosmic rays incoming to the Earth at energies from $5\cdot10^{19}$\,eV to $2\cdot10^{20}$\,eV. In this range the main processes by which UHECR particles lose energy in propagating through the space are 1) disintegration of nuclei due to interactions with infrared (IR) and cosmic microwave background (CMB) photons, and 2) pion production by protons on CMB radiation. In~addition, to estimate the contribution of nuclei to secondary protons, we extend the energy range up to Lorentz factor $\gamma\!\approx\!2.1\!\cdot\!10^{11}$, so photopion production by the nuclei is also taken into account. It should be noted that the pair production process is ignored here in order to take advantage of the Lorentz factor conservation in photodisintegration and photopion production (see Eq.\,\eqref{dNdx}). Such approximation is reasonable for light nuclei but it may not be reliable for heavy species. In fact, the absorption of iron nuclei due to pair production constitutes an~essential part ($\sim\!50$\,\%) of the total absorption at energies approximately from $10^{20}$\,eV to $1.7\!\cdot\!10^{20}$\,eV \cite{Allard_2011}. But since we consider sources of UHECR located mainly at distances below 40~Mpc and the mean free path of iron nuclei exceeds 100~Mpc at the energy range mentioned, our assumption is reasonable.

Computing the mean free paths of particles requires information about IR background radiation and cross sections of the above--mentioned processes. Data on the spectral density of IR radiation in host galaxies and intergalactic medium are taken from \cite{Porter} and \cite{Stecker}, respectively. Our calculations show that interactions of nuclei with galactic background do not lead to any substantial changes, thus the results presented below are obtained assuming only the intergalactic absorption. The cross sections of different photodisintegration channels for isotopes of nuclei from $\mathrm{Li}$ to $\mathrm{Fe}$ are computed with the TALYS code \cite{TALYS}. For nuclei $^2\mathrm{H}$, $^3\mathrm{H}$, $^3\mathrm{He}$ and $^4\mathrm{He}$, we use the total photodisintegration cross sections from the Geant4 code \cite{GEANT}. Each of them includes two reaction channels assumed as equiprobable. The photopion production cross sections for $^{56}\mathrm{Fe}$, $^4\mathrm{He}$ and $\mathrm{p}$ are also derived by means of Geant4. Corresponding cross sections for nuclei with $A\!>\!4$ are determined by multiplying that for $^{56}\mathrm{Fe}$ by $A/56$ as it follows from the concept of a~universal curve \cite{Nedorezov,MacCornick}. In the case of $A\!\leqslant\!3$ the cross sections are obtained by interpolation between those for $^4\mathrm{He}$ and $\mathrm{p}$. A~comparison between mean free paths calculated for proton and iron nucleus and those from \cite{Kampert} shows that they are in reasonable agreement at energies under consideration.

\section{Calculations}
In this section we present computational approaches used to describe propagation of protons and heavy nuclei as well as to construct their spectra.

\subsection{Nuclei}\label{nuclei}
We consider both stable and radioactive nuclei (with lifetimes of more than several minutes) arising due to photodisintegration of particles accelerated at their sources (see Section \ref{assump}). Over a~hundred isotopes in total are taken into account.

When an~ultrahigh--energy heavy nucleus propagates through the space, its energy per nucleon remains constant and its total energy is shared between particles in proportion to their masses after disintegration. Thus Lorentz factor $\gamma$ is the same for all nuclei produced. This is not correct for photopion production because pions take some energy away. But this process becomes dominant at very high energies when one needs to estimate the flux of secondary protons. Since the number of nuclei with such energies within the sources is exceedingly small and, as indicated below, their contribution to the secondary protons can be neglected, we assume that the energy is shared only between a~nucleon and a~fragment in proportion to their masses. Note that only one pion production is considered. Two pions production begins to contribute substantially at energies $\gtrsim\!550$~MeV in the mirror system (in which the nucleus is at rest), so this assumption is quite reasonable for interactions between nuclei with $\gamma\!\lesssim\!1.6\!\cdot\!10^{11}$ and CMB photons. Lastly, we presume that protons and neutrons are produced equiprobably.

Our problem can be formulated as a~chain of differential equations for the secondary nuclei resulting from an~initial one. All nonhomogeneous terms of any equation at a~given argument are defined if previous equations of the chain are already solved. Let $N_A^Z(x)$ be the number of nuclei with charge~$Z$ and mass number~$A$ at a~distance $x$ and $\lambda^i$ be the reciprocal of the mean free path for a~reaction channel $i$. We consider the following channels: photodisintegration with emitting one or more nucleons, $\alpha$-particle, etc. (every isotope accounts for eight channels of photodisintegration on average), beta--decay and pion production. Then change of the~nuclei flux with distance is described by the differential equation:
\begin{multline}\label{dNdx}
\frac{\,dN^A_Z}{dx}=-N^A_Z(x)\,\lambda^\mathrm{tot}_{(A,Z)}+\\
\sum\limits_{k,l}N^{A+k}_{Z+l}(x)\sum\limits_{m}\lambda^{m}_{(A+k,Z+l)}
+N^A_{Z+1}(x)\,\lambda^{\beta^+}_{(A,Z+1)}+\\
N^A_{Z-1}(x)\,\lambda^{\beta^-}_{(A,Z-1)}+
\!\!\!\sum\limits_{n=0,\,1}\!\!\!N^{A+1}_{Z+n}(x)\,\frac{\,\lambda^{\pi}_{(A+1,Z+n)}}{2},
\end{multline}
where the~reciprocal of the~total mean free path takes on form
\begin{equation}
\lambda^\mathrm{tot}=\sum\limits_{i}\lambda^{i}+\lambda^{\beta^+}\!\!+
\lambda^{\beta^-}\!\!+\lambda^{\pi}.
\end{equation}

The expression for $\lambda^i$ in the case of a~nucleus propagating through isotropic background with differential density $n(\varepsilon)$, is given by\,\cite{Stecker_1968,Berezinsky_1990}:
\begin{equation}\label{lam_i}
\lambda^i=\frac{1}{2\gamma^2}\!\int\limits_{\varepsilon^\prime_{th}}^{\,\infty}\!
d\varepsilon'\sigma_i(\varepsilon')\,\varepsilon'
\!\!\!\!\int\limits_{\varepsilon'\!/2\gamma}^{\,\infty}\!\!\!
d\varepsilon\,\frac{\,n(\varepsilon)}{\varepsilon^2}\,,
\end{equation}
where $\varepsilon^\prime_{th}$ is the threshold energy in the~mirror system and $\sigma_i$ is the cross section of the channel~$i$.

The solution of Eq.\,\eqref{dNdx} can be represented by:
\begin{multline}
N^A_Z(x_{n+1})=N^A_Z(x_{n-1})\,e^{-\lambda^\mathrm{tot}_{(A,Z)}(x_{n+1}-x_{n-1})}\,+\\
\sum\limits_{i,j}\lambda_{(A+i,Z+j)}\!\!\!\!\int\limits_{x_{n-1}}^{\,x_{n+1}}\!\!\!\!
N^{A+i}_{Z+j}(x')\,e^{-\lambda^\mathrm{tot}_{(A,Z)}(x_{n+1}-x')}dx',
\end{multline}
where integrals in the last term are calculated with Simpson's rule. This algorithm was used earlier in \cite{Kalmykov}. As a~control, we determine the total number of nucleons, which must be equal to the mass number of an~initial nucleus, at each integration step $x_{n}$. Accuracy of the test is better than $10^{-4}$ for all distances of interest.

We should emphasize that the solution procedure becomes more complicated if one includes also the energy losses due to pair production as the energy per nucleon does not conserve in this case.

\subsection{Protons}
The dominant process at the energies in question, which affects protons during their propagation through intergalactic medium, is photoproduction of pions. For the reason stated above we consider production of only one pion. In this case the inelasticity coefficient can be derived analytically, given that the initial energy $E_\mathrm{p}$ of a proton in the laboratory system and the energy $\varepsilon'$ of a photon in the mirror system are known.

At given $E_\mathrm{p}$ a value of $\varepsilon'$ varies with both the photon energy $\varepsilon$ and the proton--photon collision angle in the laboratory system. Values of this angle are simulated from isotropic distribution. The distribution of $\varepsilon'$ follows from Eq.\,\eqref{lam_i}:
\begin{equation}
f(\varepsilon')\,d\varepsilon'=\sigma(\varepsilon')\,\varepsilon'd\varepsilon'
\!\!\!\!\int\limits_{\varepsilon'\!/2\gamma_\mathrm{p}}^{\,\infty}\!\!\!
d\varepsilon\,\frac{\,n(\varepsilon)}{\varepsilon^2}\,.
\end{equation}

After specifying the photon energy $\varepsilon'$, the energy and momentum of a~nucleon produced are calculated in the center--of--mass system. Further they are used to determine the corresponding quantities in the laboratory system and to calculate the inelasticity coefficient. All formulae required can be found, for instance, in~\cite{Byckling}.

Thus applying Monte--Carlo simulation we can obtain a function $W_\mathrm{p}(E_\mathrm{p},E,X)dE$ which represents the probability that a proton with initial energy $E_\mathrm{p}$ is localized in the range of $(E,E\!+\!dE)$ after travelling a distance $X$, and satisfies the normalization condition:
\begin{equation}
\int\limits_{0}^{\;\;E_\mathrm{p}}\!W_\mathrm{p}(E_\mathrm{p},E,X)dE=1.
\end{equation}
The number of trial histories in our calculations is equal to $10^5$.

\begin{figure}[t]
\begin{center}
\includegraphics[height=17pc]{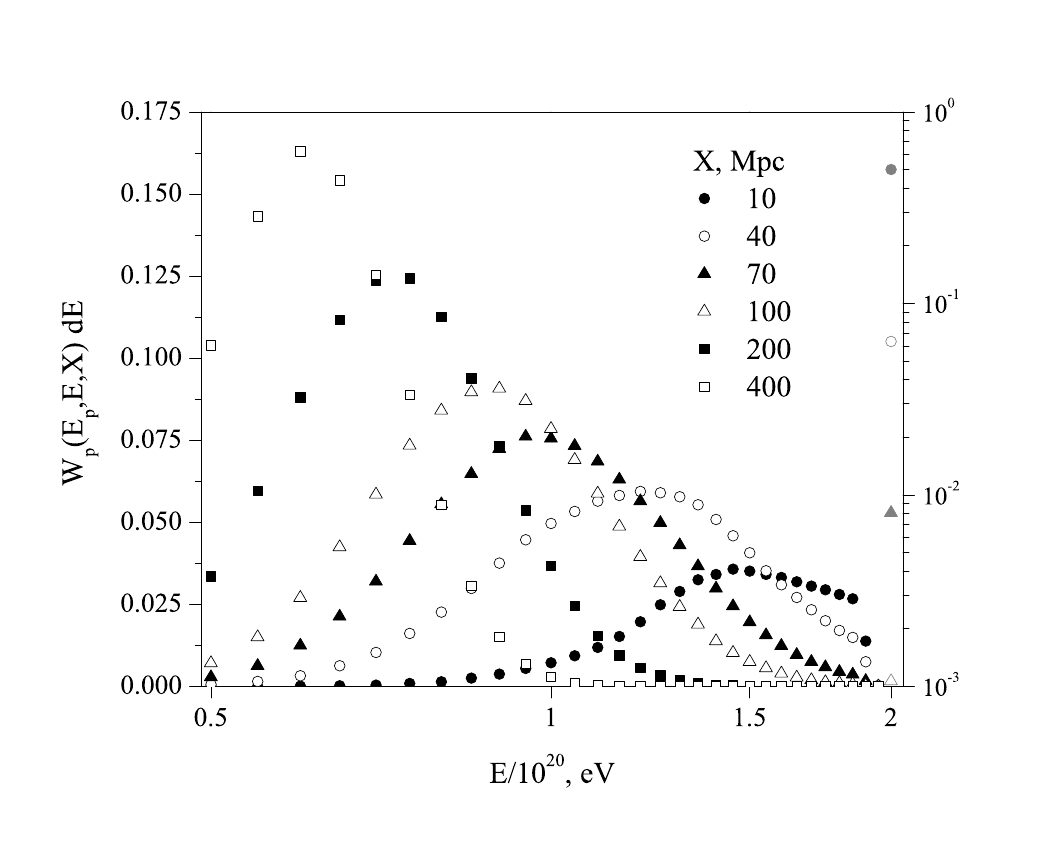}
\end{center}
\vspace{-1.5pc}
\caption{Proton probability function $W_\mathrm{p}(E_\mathrm{p},E,X)dE$ at different travelled distances $X$, with $E_\mathrm{p}\!=\!2\!\cdot\!10^{20}$\,eV and $dE\!=\!5\!\cdot\!10^{18}$\,eV. The values of points at energy $E_\mathrm{p}$ are indicated on the right scale and give the probability that a~proton conserves its initial energy after propagating a distance $X$.}\label{wprot}
\end{figure}

Figure~\ref{wprot} illustrates the probability function at $E_\mathrm{p}\!=2\!\cdot\!10^{20}$\,eV and different distances. The curves presented are smooth whereas they have pronounced peaks if mean values of the inelasticity coefficient are used.

\subsection{Spectra}
The calculation scheme presented above allows us to construct spectra of both primary (i.e. accelerated at sources) and secondary (i.e. generated as a result of photodisintegration and photoproduction of heavy nuclei) protons as well as spectra of nuclei at different distances from sources.

First we consider the calculation procedure of the spectra produced by protons. The intensity of primary ones with energy $E$ at a distance $X$ can be derived from
\begin{equation}
I_\mathrm{p}^{(1)}(E,X)=z_\mathrm{p}\!\int\limits_E^{\;\infty}C_\mathrm{p}(E')
W_\mathrm{p}(E'\!,E,X)\,dE',
\end{equation}
where $C_\mathrm{p}$ is given by
\begin{equation}\label{C1}
C_\mathrm{p}(E')=(E'\!/E_\mathrm{min})^{-\gamma_0}
\end{equation}
for energies $E'\!\leqslant\!E^\mathrm{p}_\mathrm{max}$, and
\begin{equation}\label{C2}
C_\mathrm{p}(E')=(E^\mathrm{p}_\mathrm{max}/E_\mathrm{min})^{-\gamma_0}
(E'\!/E^\mathrm{p}_\mathrm{max})^{-(\gamma_0+4)}
\end{equation}
otherwise, with $E_\mathrm{min}\!=5\!\cdot\!10^{19}$\,eV. This factor sets the number of particles in the injection spectrum at a given energy.

In the case of secondary protons the so-called source function is required:
\begin{equation}
S_\mathrm{p}^n(E'\!,x)=dN_\mathrm{p}^n(E'\!,x)/dx,
\end{equation}
where $N_\mathrm{p}^n(E'\!,x)$ yields the amount of the protons with energy $E'$, produced in travelling a distance $x$ by an initial nucleus $n$ with mass number $A_n$. Then their intensity is described~by
\begin{equation}
I_\mathrm{p}^n(E,X)=\!\int\limits_{\!0}^{\;X}\!dx\:\times
\int\limits_{E}^{\;\infty}\!C_n(E')\,S_\mathrm{p}^n(E'\!,x)\,W_\mathrm{p}(E'\!,E,X\!-\!x)\,dE'
\end{equation}
and the total intensity is the sum over all nuclei accelerated:
\begin{equation}
I_\mathrm{p}^{(2)}(E,X)=\!\sum\limits_nz_nA_n^{\gamma_0-1}I_\mathrm{p}^n(E,X)\Bigl/\sum\limits_nz_nA_n^{\gamma_0-1}.
\end{equation}
Since the energy of the initial nucleus is $A_n$ times as great as that of the nucleons produced, the factor $C_n$ can be written as
\begin{equation}\label{Cn1}
C_n(E')=(A_nE'\!/E_\mathrm{min})^{-\gamma_0}
\end{equation}
if $A_nE'\!\leqslant\!E^n_\mathrm{max}$, and
\begin{equation}\label{Cn2}
C_n(E')=(E^n_\mathrm{max}/E_\mathrm{min})^{-\gamma_0}
(A_nE'\!/E^n_\mathrm{max})^{-(\gamma_0+4)}
\end{equation}
otherwise.

Nuclei spectra can be constructed in a~similar way. It is conventional in cosmic rays physics to consider groups of nuclei. In our calculations we also combine them in several groups depending on their mass number. These are the helium group $\mathrm{He}$ ($A\!=\!3-4$, $\langle A_\mathrm{He}\rangle\!=\!4$) and the groups $\mathrm{L}$, $\mathrm{M}$, $\mathrm{H}$, $\mathrm{VH}$ of light ($A\!=\!5\!-\!11$, $\langle A_\mathrm{L}\rangle\!=\!8$), mean ($A\!=\!12\!-\!19$, $\langle A_\mathrm{M}\rangle\!=\!16$), heavy ($A\!=\!20\!-\!39$, $\langle A_\mathrm{H}\rangle\!=\!32$) and very heavy ($A\!=\!40\!-\!56$, $\langle A_\mathrm{VH}\rangle\!=\!48$) nuclei, respectively. Here $\langle A\rangle$ is the mean mass number. In fact each group at a~given energy $E\!=\!\gamma\langle A\rangle m_\mathrm{p}$ contains nuclei with energies in the range of $\gamma\underline{A}m_\mathrm{p}\!\leqslant\!E\!\leqslant\!\gamma\,\overline{\!A}m_\mathrm{p}$, where $\underline{A}$ and $\overline{\!A}$ are the minimal and maximum mass numbers of the group. Nevertheless this procedure is quite justified as there exist substantial uncertainties in measuring energy and type of incoming UHECR particles. Then the intensity of a~group $g$ is defined as
\begin{equation}
I_g(E,X)=\sum\limits_nz_nA_n^{\gamma_0-1}C_n^g(E)N_{n}(E,X)\Bigl/
\sum\limits_nz_nA_n^{\gamma_0-1},
\end{equation}
where $N_n(E,X)$ is the sum of values $N^A_Z(X)$ (calculated according to Eq.\,\eqref{dNdx}) for the secondary nuclei with energy $E$ produced by an~initial particle $n$ at a~distance $X$ from its source and the index $n$ runs over all nuclei contributing to the group. For example, the group $\mathrm{VH}$ of very heavy nuclei includes particles which could be generated during propagation of $^{40}\mathrm{Ar}$, $^{40}\mathrm{Ca}$, $^{52}\mathrm{Cr}$ and $^{56}\mathrm{Fe}$. For initial nucleus $^{52}\mathrm{Cr}$ the value $N_n$ consists of all terms between $N^{40}_{18}$ and $N^{52}_{24}$. The factor $C_n^g$ depends on the mean mass number $\langle A_g\rangle$ of the group $g$ and is defined similarly to \eqref{Cn1}\,--\,\eqref{Cn2} with replacement of $A_nE'$ by $A_nE'/\langle A_g\rangle$.

The~spectra obtained in this way are expressed in relative terms, with the total number of injected particles at $E_\mathrm{min}\!=5\!\cdot\!10^{19}$\,eV taken as unity.

\section{Spectra and mass composition of UHECRs from a single source}\label{sect_single}
\begin{figure}[t]
\begin{center}
\includegraphics[height=24pc]{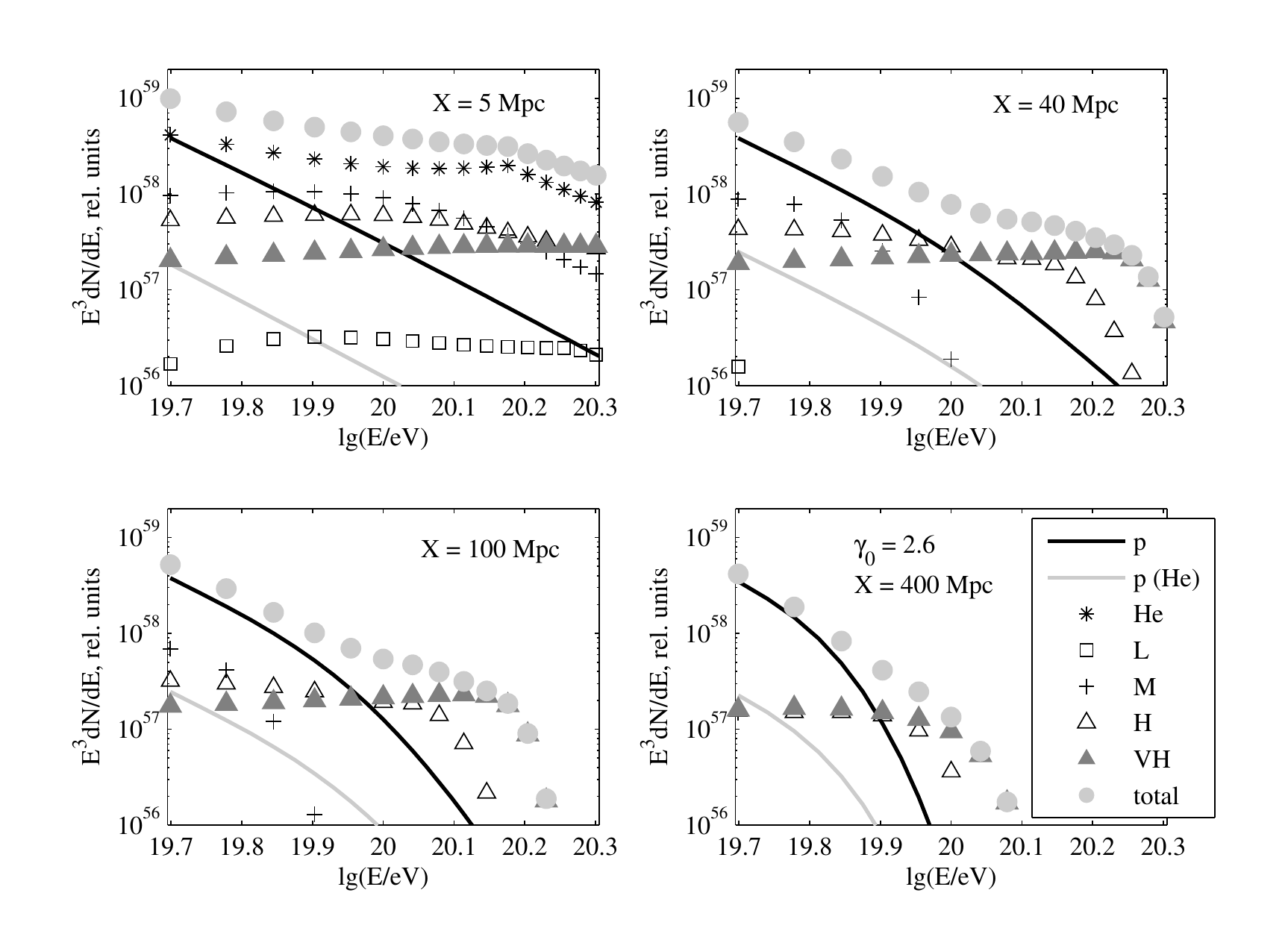}
\end{center}
\vspace{-1.5pc}
\caption{Differential energy spectra of the nuclei groups, produced by a~single source at distances $X\!=5$, 40, 100 and 400~Mpc. The flux of secondary protons generated due to disintegration of the~helium group is marked by $\mathrm{p}\,(\mathrm{He})$. The total spectrum is presented by grey circles. The number of all particles with energy $E\!=\!5\!\cdot\!10^{19}$\,eV at the source is normalized to unity. The generation spectra are described by a~broken inverse power law with index $\gamma_0\!=\!2.6$.}\label{sp26}
\end{figure}

\begin{figure}[t]
\begin{center}
\includegraphics[height=24pc]{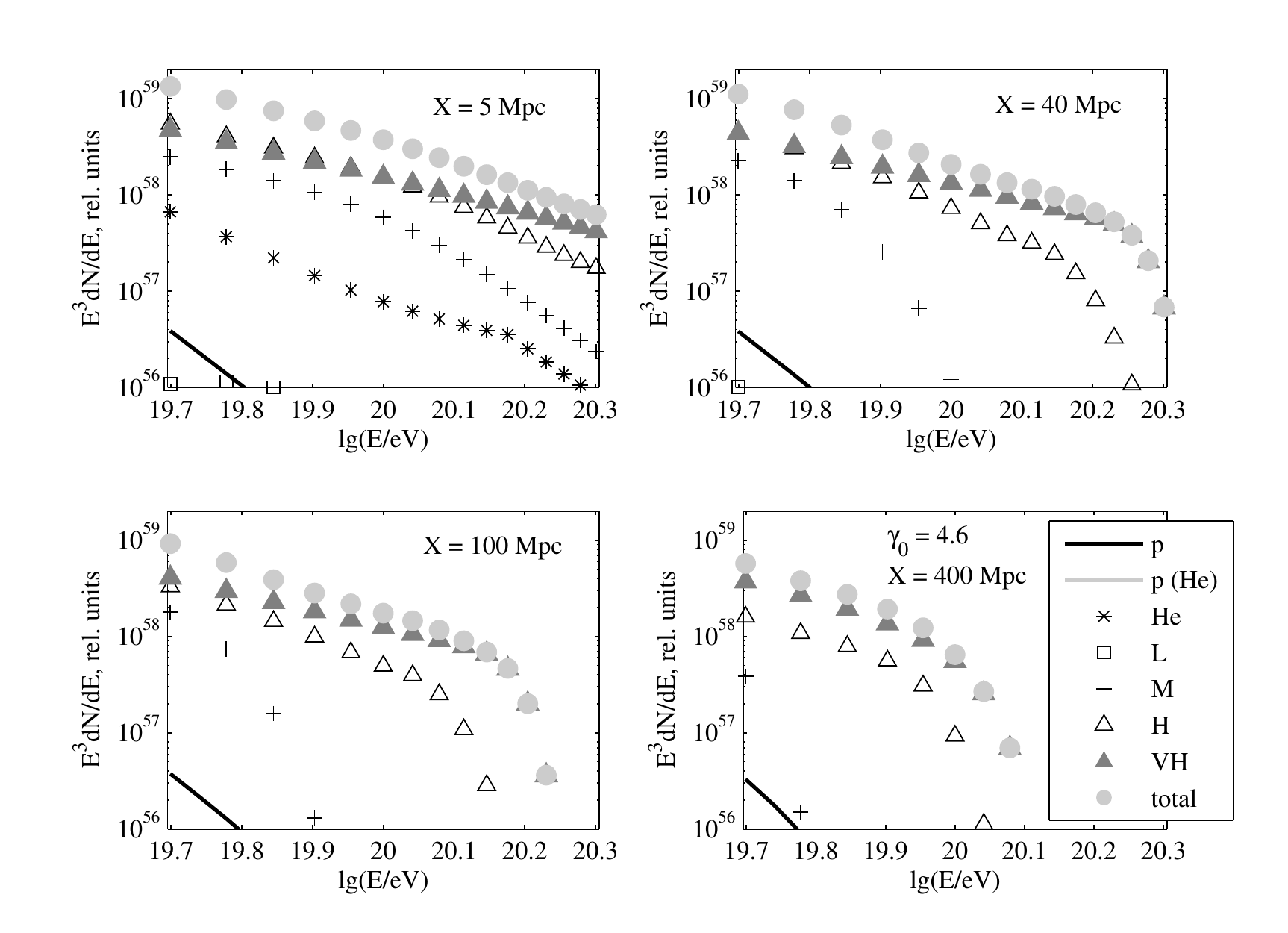}
\end{center}
\vspace{-1.5pc}
\caption{The same as Figure~\ref{sp26}, but $\gamma_0\!=\!4.6$.}\label{sp46}
\end{figure}

In this section we present spectra and mass composition of cosmic rays in the range of energies from $5\!\cdot\!10^{19}$\,eV to $2\!\cdot\!10^{20}$\,eV, produced by a~single source located at various distances~$X$. Though the CR spectrum is modified during propagation of the particles through the space, its shape at the observation level strongly depends on the injection index $\gamma_0$. Figures \ref{sp26}\,--\,\ref{sp46} illustrate the differential spectra of protons and nuclei groups, produced by the source with $\gamma_0\!=\!2.6$, 3.6, 4.6 at $X\!=5$, 40, 100, 400~Mpc. The maximum acceleration energies of the particles are evaluated from Eq.\,\eqref{maxen_nucl} and \eqref{maxen_prot}. The total flux at $E_\mathrm{min}$ is less than unity, which results from transition of more energetic protons to lower energies and disintegration of nuclei into fragments during propagation.

\begin{figure}[h!]
\begin{center}
\includegraphics[height=24pc]{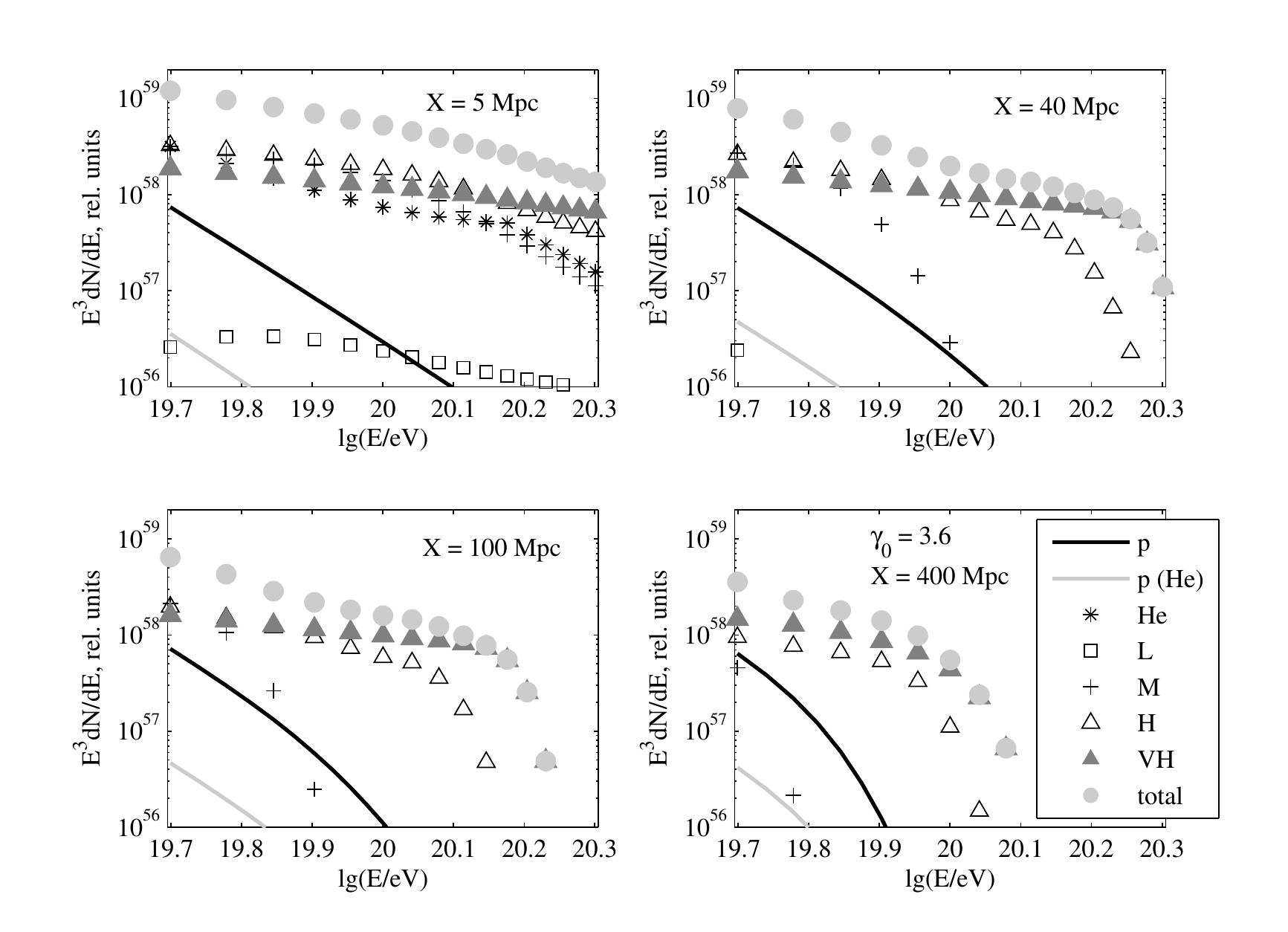}
\end{center}
\vspace{-1.5pc}
\caption{The same as Figure~\ref{sp26}, but $\gamma_0\!=\!3.6$.}\label{sp36}
\end{figure}

All these pictures have some common features. Firstly, the helium group decays rapidly since the mean free path $\lambda_\mathrm{tot}$ of these particles is small enough. Besides, there is no essential replenishment from heavier nuclei. A~noticeable flattening in the spectrum of the $\mathrm{He}$ group below $E_\mathrm{max}^\mathrm{He}$ arises from the change of dominance between two interaction mechanisms: photodisintegration and photopion production, accompanied by a~slight increase in $\lambda_\mathrm{tot}$. Secondly, although nuclei of the $\mathrm{L}$ group ($\mathrm{Li}$, $\mathrm{Be}$, $\mathrm{B}$) are absent at the source, they appear in small numbers in propagating of initial nuclei. Thirdly, as discussed above, one can neglect the~contribution of heavier nuclei to the proton flux. The highest proton yield is supported by the $\mathrm{He}$ group, as a~consequence of its fast decrease. As can be seen, the fluxes of all nuclei groups undergo the cutoff at sufficiently large distances from the source. This effect should be more pronounced if their absorption due to pair production is taken into account.

\begin{figure}[h!]
\begin{center}
\includegraphics[height=12pc]{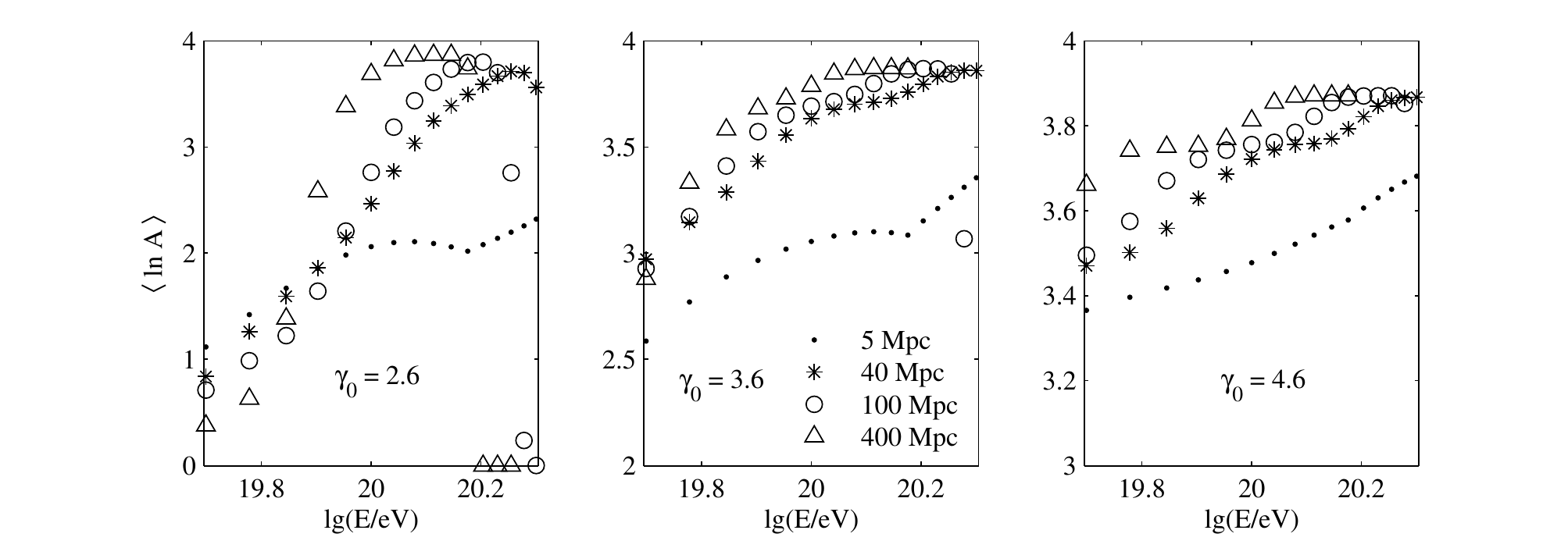}
\end{center}
\vspace{-1pc}
\caption{Energy dependence of the~mean logarithmic mass number of particles composing the~total spectra from Figures~\ref{sp26}\,--\,\ref{sp46}.}\label{comp}
\end{figure}

At a~given value of the injection index $\gamma_0$ it is easy to estimate the CR composition. Figure~\ref{comp} demonstrates the energy dependence of the mean logarithmic mass of particles composing the aforementioned total spectra:
\begin{equation}
\langle\ln A\rangle=\sum\limits_{g}I_g(E)\ln\langle A_g\rangle\!\Bigl/\,\sum\limits_{g}I_g(E),
\end{equation}
where $g$ runs over all nuclei groups, including protons. As can be seen, the larger is $\gamma_0$, the heavier is the UHECR composition, especially at energies below $10^{20}$\,eV.

\section{Comparison with the HiRes and Auger data}\label{res}
When calculating spectra, we make a~simple assumption that all nearby sources are identical in cosmic--ray intensity as well as in composition and acceleration conditions for particles. Therefore the contribution of a~single source located at a~distance $X$ to the total flux should be taken into account with weight $X^{-2}$. Then the average distance to the nearby sources is close to 9~Mpc. In this case the UHECR spectrum is modified only slightly and resembles the result obtained for a~single source located at a~distance of 5~Mpc.

\begin{figure}[h!]
\begin{center}
\includegraphics[height=20pc]{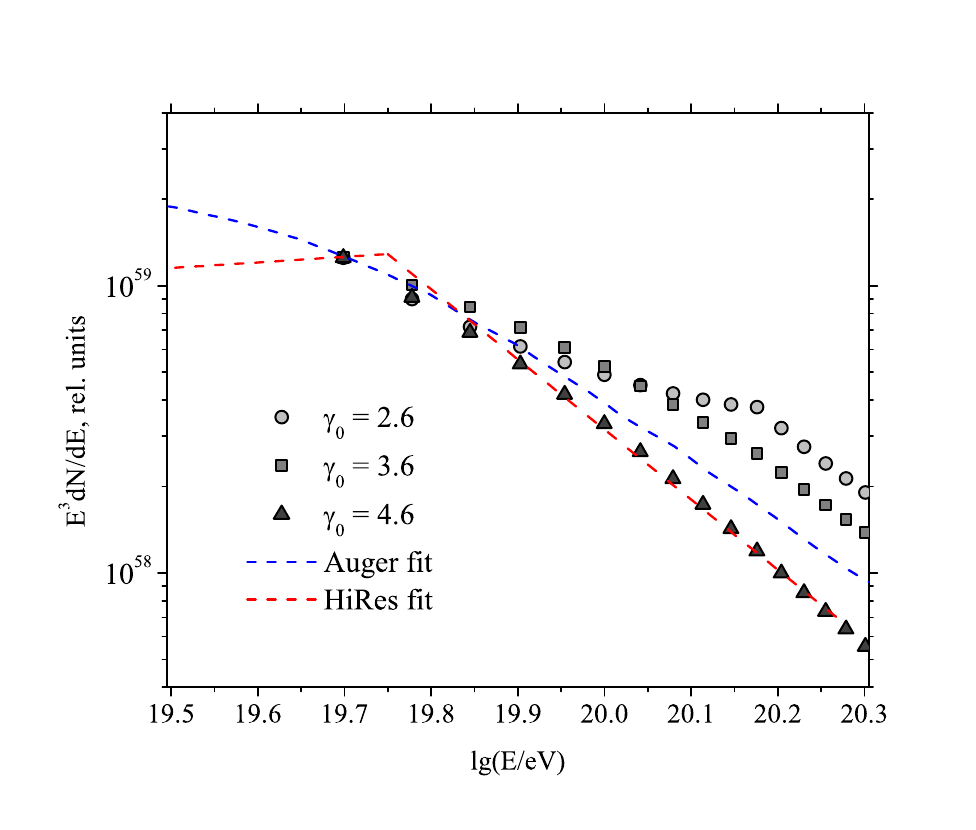}
\end{center}
\vspace{-2pc}
\caption{Total energy spectra of particles, produced by nearby sources, with injection indices $\gamma_0\!=\!2.6$, 3.6, and 4.6. Values of the maximum acceleration energies are evaluated from Eq.\,\eqref{maxen_nucl}, \eqref{maxen_prot}. The~HiRes and Auger fits are shown for comparison. All data are normalized to unity at $5\!\cdot\!10^{19}$\,eV.}\label{sp_comparison}
\end{figure}

Figure~\ref{sp_comparison} demonstrates the total spectra produced by the nearby sources, with injection indices $\gamma_0\!=\!2.6$, 3.6, and 4.6. The maximum acceleration energies of the particles are evaluated from Eq.\,\eqref{maxen_nucl} and \eqref{maxen_prot}. The fits to experimental data, provided by HiRes \cite{Sokolsky_2009} and Auger \cite{PAO_2009}, are also presented. All data are normalized to unity at energy $5\!\cdot\!10^{19}$\,eV. A~comparison with the HiRes broken power law fit shows that there is no break in the spectrum at $\gamma_0\!=\!2.6$, which is the most appropriate for light composition. In~addition, the spectra at $\gamma_0\!=\!2.6$ and 3.6 do not reveal sufficient suppression. It is evident that in the case of short distances the suppression of the UHECR spectrum can arise rather due to acceleration limits at the sources than as a~result of interactions of CR particles with intergalactic background photons. Hence the spectral break at the GZK threshold $E_{GZK}$ may be attained if we take it as the maximum acceleration energy of some nuclei group. The group, of course, should be dominant at energies below $E_{GZK}$ to keep the break in the total spectrum at the observation level. According to the HiRes results \cite{Sokolsky_2009}, the GZK cutoff starts from approximately $5.7\!\cdot\!10^{19}$\,eV and the broken power law fit yields indices of roughly 2.8 and 5.4 below and above it, respectively. The Auger spectrum also testifies for the CR flux suppression but its fit has a~smooth shape \cite{PAO_2009,PAO_2011}. In order to reproduce the spectral break and light composition simultaneously in the calculated spectra at lower $\gamma_0$, we set the maximum acceleration energy for protons equal to $E_{GZK}\!=\!5.7\!\cdot\!10^{19}$\,eV. The values of $E_\mathrm{max}$ for the rest of initial nuclei, calculated from Eq.\,\eqref{maxen_nucl}, are also increased by a~factor of~1.4, however the $Z^{2/3}$--dependence is retained. Below all results are obtained with the assumption of these new maximum acceleration energies.

\subsection{Spectra}
\begin{figure}[t]
\begin{center}
\includegraphics[height=20pc]{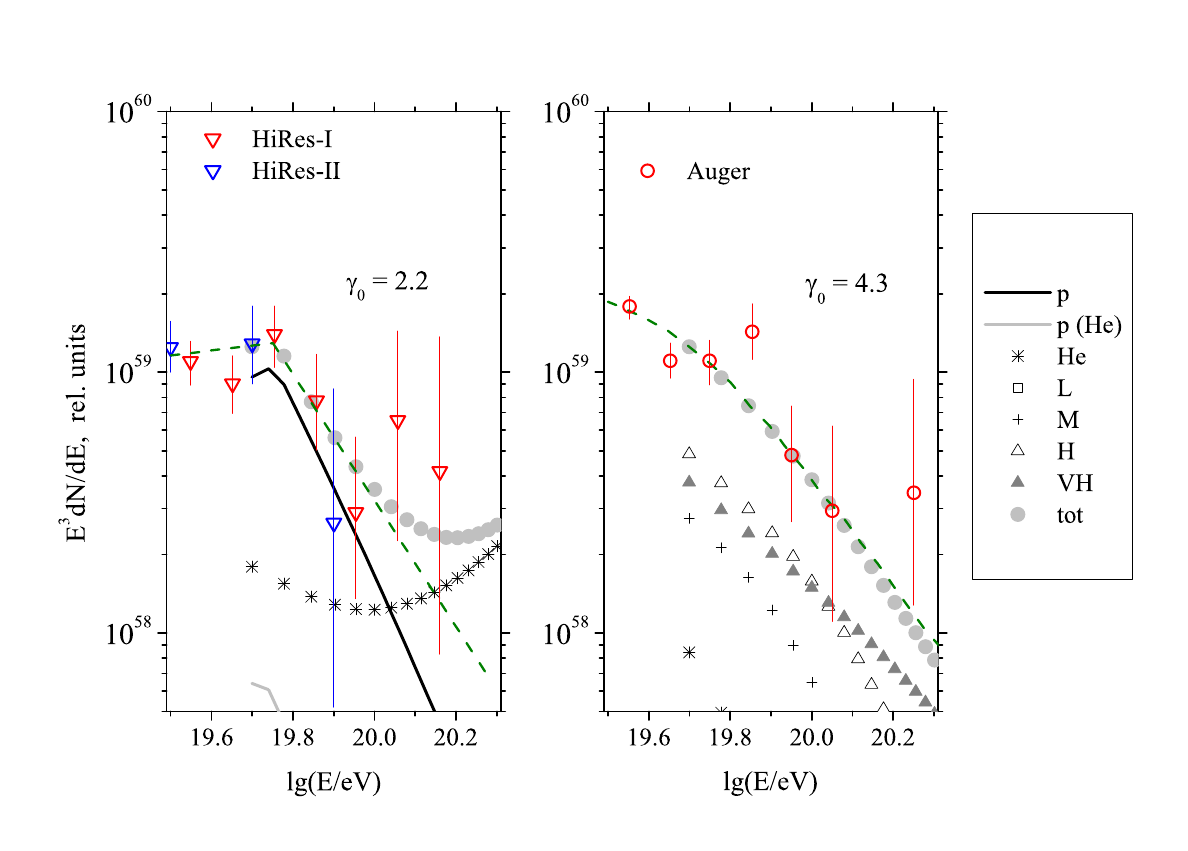}
\end{center}
\vspace{-1.5pc}
\caption{Differential energy spectra of the nuclei groups, produced by nearby sources with injection indices $\gamma_0\!=\!2.2$ (the left panel) and 4.3 (the right panel). The flux of secondary protons generated by the helium group is marked by $\mathrm{p}\,(\mathrm{He})$. The total spectrum is presented by grey circles. The HiRes and Auger experimental points along with their fits (grin dashed curves) are shown for comparison on the left and right panels, respectively. All data are normalized to unity at $5\!\cdot\!10^{19}$ eV.}\label{source_sp}
\end{figure}

Our calculations show that optimal values of the injection index $\gamma_0$ for representing the HiRes and Auger data are 2.2 and 4.3, respectively. In~Figure~\ref{source_sp} the calculated spectra produced by nearby sources are compared with experimental points and their fits. All data are normalized to unity at $5\!\cdot\!10^{19}$\,eV. The energy range is extended down to $10^{19.5}$\,eV to show the behavior of the measurements at lower energies. According to the left panel, our results are described well enough by a~broken power law with indices of 2.8 and 5.4 below and above the GZK--cutoff energy at $10^{19.75}$\,eV, as predicted by HiRes. This agreement is valid up to the energy $\approx\!10^{20}$\,eV whereas at higher energies the above--mentioned feature of the helium group affects the calculated spectrum strongly, making it flatten (see Section~\ref{sect_single}). In contrast, the right panel demonstrates a~good accord between the spectrum with $\gamma_0\!=4.3$ and the smooth function given by Auger, in all the energy range under study. Computational results presented in \cite{PAO_2009} produce this smooth function within a~hypothetical model with a~pure iron composition. In our model the dominance of heavy nuclei results from a~large value of the injection index.

\subsection{Mass composition}
In this subsection we discuss the UHECR composition. At present this topic is one of the most attractive problems in high--energy astrophysics. In fact, the largest facilities, HiRes and Pierre Auger Observatory, yield conflicting data. The~Auger measurements of the mean shower maximum $\langle X_\mathrm{max}\rangle$ and its root--mean--square fluctuations show that mass composition of extragalactic cosmic rays grows heavy with energy \cite{Auger_xmax}. On the contrary, the HiRes analysis of the $\langle X_\mathrm{max}\rangle$ fluctuations indicates the proton dominance \cite{HiRes_xmax}. Though both compositions are obtained for particles with energies below $5\!\cdot\!10^{19}$\,eV, we extend these results to higher energies, assuming that there is no significant change in them.

\begin{figure}[t]
\begin{center}
\includegraphics[height=12pc]{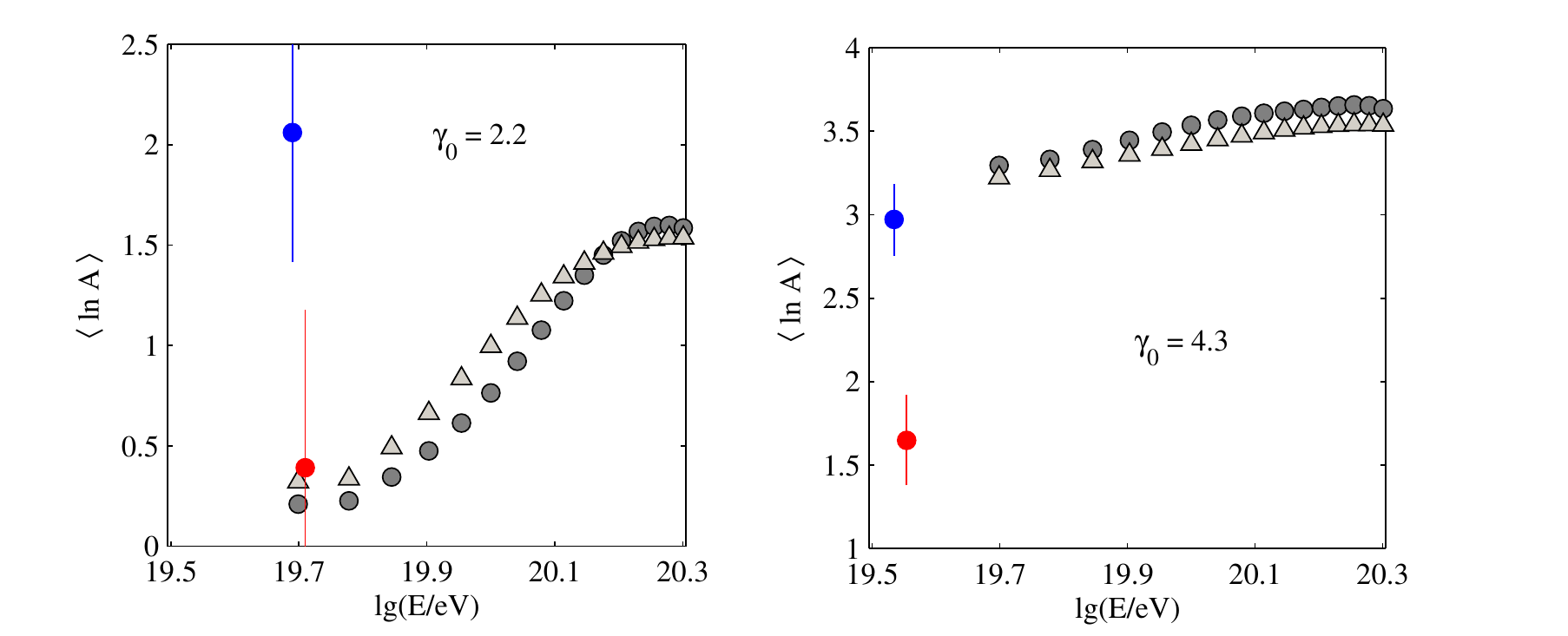}
\end{center}
\vspace{-1pc}
\caption{Energy dependence of the mean logarithmic mass number of particles composing the total spectra from Figures~\ref{source_sp} (light grey triangles) and \ref{uniform_sp} (dark grey circles). The results extracted from the highest energy experimental values of $\langle X_\mathrm{max}\rangle$ provided by HiRes (the left panel) and Auger (the right panel) are shown for comparison. The points calculated from the EPOSv1.99 (blue) and QGSJET01 (red) predictions are shifted by $\Delta\lg E\!=\!0.01$ apart from their initial energy.}\label{disunif_comp}
\end{figure}

Figure~\ref{disunif_comp} shows the calculated energy dependence of the mean logarithmic mass number of particles composing the total spectra from Figures~\ref{source_sp} and \ref{uniform_sp}. For comparison, the values of $\langle\ln A\rangle$ are also retrieved from the mean shower maximum data measured at the highest energy $E_c$ by HiRes \cite{HiRes_xmax} ($E_c\!\approx\!5\!\cdot\!10^{19}$\,eV) and Auger \cite{Auger_xmax} ($E_c\!\approx\!3.5\!\cdot\!10^{19}$\,eV), by interpolation between the model predictions for protons and iron nuclei. We chose two models which provide extreme results: EPOSv1.99 and QGSJET01 (see Figure\,3 in\,\cite{Auger_xmax}). The color points are shifted relative to $E_c$ by $\Delta\lg E\!=\!0.01$ for clarity. There is no possibility to describe the experimental data of both facilities in the framework of one model. Hopefully, this discrepancy will be overcome after the models used for the comparison undergo necessary modifications in response to the LHC data. However, if we compare experimental values of $\langle X_\mathrm{max}\rangle$, it is evident that the HiRes composition is lighter than the Auger one at ultrahigh energies. According to our calculations, the composition at $\gamma_0=2.2$ seems to be a~mixture of protons and light nuclei such as helium, up to $7\!\cdot\!10^{19}$\,eV, which is consistent with the HiRes data. At higher energies it grows gradually heavier. In contrast, the composition at $\gamma_0\!=\!4.3$ is compatible with the Auger results. Thus one can conclude tentatively that mass composition measurements make it possible to distinguish between different values of the injection index. Though, at the moment several models of hadron interactions are used and their predictions concerning the shower maximum depth dependence on energy differ significantly.

\section{Discussion}
The necessity of nearby sources of UHECR nuclei has been discussed during the last few years. As it was mentioned above, Seyfert galaxies located at distances of $\lesssim\!40$~Mpc were suggested as probable sources of UHECRs \cite{Uryson_AR}. The correlation between the arrival directions of ultrahigh--energy events and AGNs within 75~Mpc was found in the Auger data \cite{Auger_2007,Abreu_2010}. Specific quantitative constraints on the UHECR source population were obtained in~\cite{Taylor_2011}. For uniformly distributed sources with pure silicon or iron composition and intermediate cutoff energies ($10^{20.5}\!-\!10^{21}$\,eV for iron nuclei), the authors found that the lack of nearby sources makes impossible to get a~good agreement with the Auger data and that the nearest sources have to be within 60~Mpc and 80~Mpc for silicon and iron sources, respectively.

\begin{figure}[t]
\begin{center}
\includegraphics[height=20pc]{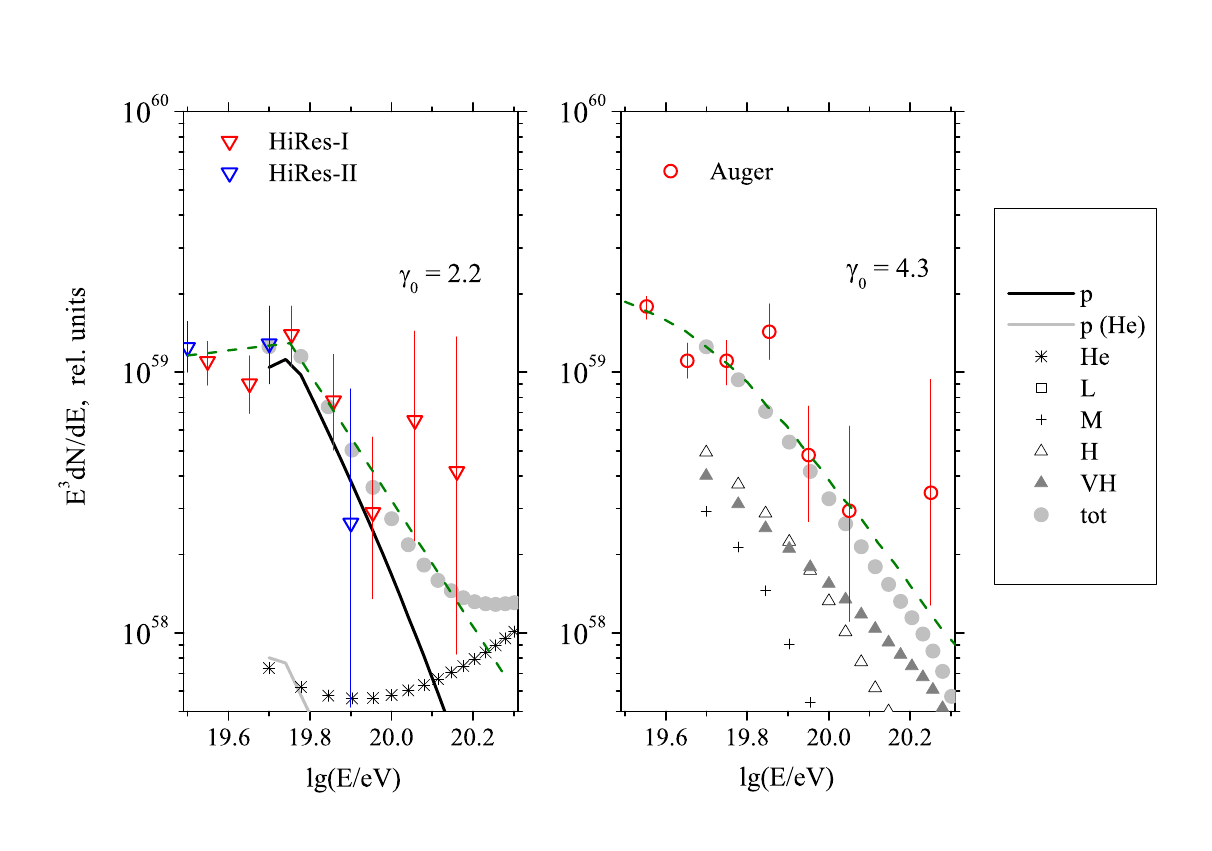}
\end{center}
\vspace{-1.5pc}
\caption{Differential energy spectra of the nuclei groups, produced by sources uniformly distributed within a~layer from 1~Mpc to 40~Mpc, with injection indices $\gamma_0\!=\!2.2$ (the left panel) and 4.3 (the right panel). The description of this figure is the same as that of Figure~\ref{source_sp}.}\label{uniform_sp}
\end{figure}

In our work we considered Seyfert galaxies at distances of $\lesssim\!40$~Mpc as sources of UHECRs and found that it is possible to describe both HiRes and Auger measurements, varying the injection index~$\gamma_0$. Moreover, diverting from the specific type of sources, we calculated the energy spectra and mass composition of particles produced by sources uniformly distributed within a~spherical layer from 1~Mpc to 40~Mpc, with all other factors being the same. As can be seen from Figures~\ref{disunif_comp} and \ref{uniform_sp}, there is also an agreement with the experimental data provided by the facilities, so the results obtained are not too sensitive to the distribution form of the nearby sources.

It is worthwhile mentioning that our calculations are performed in the framework of the widespread notion that the influence of extragalactic magnetic fields (EGMF) on the propagation of UHECR particles can be neglected. The same assumption is mandatory for identifying the highest energy events recorded by HiRes and Auger with astrophysical objects. In fact, Larmor radius at $10^{20}$\,eV exceeds 400~Mpc even for iron nuclei in homogeneous magnetic fields below $10^{-11}$\,G. However, there is no generally accepted opinion concerning EGMF. The influence of random magnetic fields with $B\!=\!10^{-10}\!-\!10^{-9}$\,G and a~coherence scale of 1~Mpc was considered in \cite{Taylor_2011}. The authors found no additional constraints on the upper bound imposed on the~nearest sources distances. But if the magnetic field is significantly stronger ($10^{-8}\!-\!10^{-7}$\,G) then the UHECR sources must be located at distances of a~few~Mpc, with the only appropriate candidate being the nearest active radio galaxy Cen\,A~\cite{Piran_2010}.

\section{Conclusions}
In this paper we have analyzed the propagation of UHECR particles in the intergalactic medium from sources to the Earth and obtained their spectra and mass composition in the energy range of $(0.5\!-\!2)\!\cdot\!10^{20}$\,eV, based on the following assumptions: the UHECR sources are considered to be Seyfert galaxies located within the radius of $\approx\!40$~Mpc; mass composition of the particles at the sources resembles space content; relations between maximum energies of different nuclei are retained according to the acceleration model suggested for moderate--power sources; generation spectra obey an inverse power law with arbitrary index $\gamma_0$. The higher is the value of~$\gamma_0$, the heavier is the composition of nuclei accelerated. Over a~hundred isotopes arising due to interactions with background photons are taken into account.

Our calculations have shown that the~conjecture of nearby sources located at distances of less than 40~Mpc is compatible with the HiRes and Auger experimental data. Although these data are contradictory, it has been possible to represent them in the framework of a~unifying approach with varying injection index.

It is worth noting that our results depend crucially on the model parameters. We hope that a~larger number of UHECR events, which should be available in the near future, will allow to examine generation spectra at sources in more detail.

\end{document}